\documentclass[aps,pra,preprint,groupedaddress,showpacs]{revtex4}
\usepackage{graphicx} 
\usepackage{dcolumn}  
\usepackage{bm}       

\begin{document}
\preprint{Version 1.1}

\title{Radiative Corrections to One-Photon Decays of Hydrogenic Ions}

\author{J. Sapirstein}
\email[]{jsapirst@nd.edu}
\affiliation{Department of Physics, University of Notre Dame,
             Notre Dame, IN 46556}

\author{K. Pachucki}
\email[]{krp@fuw.edu.pl}
\affiliation{Institute of Theoretical Physics, Warsaw University,
             Hoza 69, 00-681 Warsaw, Poland}

\author{K. T. Cheng}
\email[]{ktcheng@llnl.gov}
\affiliation{University of California, Lawrence Livermore National
             Laboratory, Livermore, CA 94550}

\date{\today}

\begin{abstract}
Radiative corrections to the decay rate of $n=2$ states of hydrogenic
ions are calculated. The transitions considered are the M1 decay of the
$2s$ state to the ground state and the E1(M2) decays of the $2p_{1/2}$ and
$2p_{3/2}$ states to the ground state. The radiative corrections start
in order $\alpha (Z\alpha)^2$, but the method used sums all orders of
$Z\alpha$. The leading $\alpha (Z\alpha)^2$ correction for the E1 decays
is calculated and compared with the exact result. The extension of the
calculational method to parity nonconserving transitions in neutral atoms
is discussed.
\end{abstract}


\maketitle

\section{Introduction}

Radiative corrections to decay rates in atoms and ions have not been as
thoroughly studied as other kinds of radiative corrections, such as the
Lamb shift \cite{PJM} and corrections to hyperfine splitting (hfs)
\cite{hfs1,hfs2}. An exception is the case of the exotic atom positronium,
where differences between the lowest-order decay rates and experiment of
-0.6 and -2.2 percent are present for parapositronium \cite{paraexp} and
orthopositronium \cite{orthoexp}, respectively. In both cases the bulk
of the difference is accounted for by one-loop radiative corrections
\cite{Harris-Brown,Lepage}, which enter in order $\alpha$ with large
coefficients, and the present theoretical interest has advanced to
the level of two-loop radiative corrections \cite{CMY,AFS}.

For other hydrogenlike atoms, theoretical work on one-loop corrections to
M1 decays has been carried out in Ref.~\cite{Lin-Feinberg,Barbieri-Sucher}.
These papers established that, unlike positronium, the order $\alpha$
correction has a vanishing coefficient, but did not calculate the actual
correction, which enters in order $\alpha (Z\alpha)^2$. A calculation for
E1 decays of the $\alpha (Z\alpha)^2 \ln(Z\alpha)$ correction has been
carried out in Ref.~\cite{Ivanov}, and is in disagreement with another
calculation associated with the experimental determination of the
Lamb shift \cite{Sok}. This situation will be discussed further in
the conclusion.

It is the purpose of the present paper to calculate radiative corrections
for the hydrogen isoelectronic sequence using methods that treat the
electron propagator exactly. In addition, a perturbative calculation for
E1 decays through order $\alpha (Z\alpha)^2$ is carried out and compared
to the exact result. While of intrinsic interest, development of these
techniques should also aid in the evaluation of radiative corrections to
parity nonconserving transitions in atoms, as will be discussed in the
conclusion.

\section{\label{sec:lowest}Lowest-order calculation}

While the first calculations of the decay rate of hydrogen date back
to the beginning of quantum mechanics, fully relativistic calculations
needed for calculations of highly-charged hydrogenlike ions were first
carried out in the early 1970's \cite{Drakeh}. We briefly present the
theory here using techniques that will be extended to the radiative
correction case. We want to use the fact that a decay rate can be
related to the imaginary part of the energy through
\begin{equation}
\Gamma = -2\,\Im(E).
\end{equation}
This is the approach taken by Barbieri and Sucher \cite{Barbieri-Sucher}.
The one-photon decay rate is connected through this formula with the
self-energy of an electron in a state $v$, which will be chosen here
to be $2s_{1/2}$, $2p_{1/2}$, or $2p_{3/2}$. We define this self-energy
as $\Sigma_{vv}(\epsilon_v)$, where
\begin{equation}
\Sigma_{ml}(E) = -ie^2 \!\int\! d^3 x\, d^3 y \!\int\!
{d^n k \over (2\pi)^n}\,
{e^{i\vec k \cdot (\vec x - \vec y)} \over k^2 + i\delta}\,
\bar{\psi}_m(\vec x) \gamma_{\mu} S_F(\vec x, \vec y; E - k_0)
\gamma^{\mu} \psi_l(\vec y),
\end{equation}
and a self-mass counterterm needed to renormalize $\Sigma$ is understood
to be included. If we set $n=4$, carry out the $d^3 k$ integration, and
represent the Dirac-Coulomb propagator by a spectral decomposition,
the above can be written
\begin{equation}
\Sigma_{ml}(E) = i\alpha \!\int\! d^3 x\, d^3 y \!\int\!
{dk_0 \over 2\pi} \sum_r
{e^{i\sqrt{k_0^2 + i\delta}\, |\vec x - \vec y|} \over |\vec x - \vec y|}
{\bar{\psi}_m(\vec x) \gamma_{\mu} \psi_r(\vec x)\,
 \bar{\psi}_r(\vec y) \gamma^{\mu} \psi_l(\vec y)
 \over E - k_0 - \epsilon_r(1 - i\delta)}.
\end{equation}
If we define
\begin{equation}
g_{ijkl}(E) = \alpha \!\int\! d^3 x\, d^3 y\,
{e^{i\sqrt{E^2 + i\delta}\, |\vec x - \vec y|} \over |\vec x - \vec y|}\,
\bar{\psi}_i(\vec x) \gamma_{\mu} \psi_k(\vec x)\,
\bar{\psi}_j(\vec y) \gamma^{\mu} \psi_l(\vec y),
\end{equation}
then the self-energy can be compactly represented as
\begin{equation}
\Sigma_{vv}(\epsilon_v) = i \!\int\! {dk_0 \over 2\pi}
\sum_m {g_{vmmv}(k_0) \over \epsilon_v - k_0 - \epsilon_m(1 - i\delta)}.
\label{eq:sigma}
\end{equation}
It will be convenient below to also introduce the function
\begin{equation}
\bar{g}_{ijkl}(E) = \alpha \!\int\! d^3 x\, d^3 y\,
{\sin(E |\vec x - \vec y|) \over |\vec x - \vec y|}\,
\bar{\psi}_i(\vec x) \gamma_{\mu} \psi_k(\vec x)\,
\bar{\psi}_j(\vec y) \gamma^{\mu} \psi_l(\vec y).
\end{equation}

To carry out the numerical evaluation of the Lamb shift,
a Wick rotation with $k_0 \rightarrow i\omega$ is performed.
The resulting expression is purely real because imaginary parts present
in the $\omega = 0 - \infty$ interval cancel against other imaginary parts
in the $\omega = \text{-}\infty-0$ interval. The imaginary part of the
self-energy arises solely from the pole term, where a bound state pole
in the first quadrant present when $\epsilon_v > \epsilon_m$ is encircled
during the Wick rotation. As we are interested in decays to the ground
state, we will not consider imaginary parts of the energy arising from
$m$ being an excited state in Eq.~(\ref{eq:sigma}). We introduce the
convention that $a$ refers to the ground state when there is no dependence
on the magnetic quantum number (as is the case for the energy $\epsilon_a$
and the self-energy $\Sigma_{aa}(\epsilon_a)$), and $b$ or $c$ refers to
the ground state when there is a dependence, with a sum over $b$ or $c$
running over the two possible values ($\pm 1/2$) of the magnetic quantum
number. We also define the lowest-order decay photon energy
$\Delta E = \epsilon_v - \epsilon_a$. It is important to emphasize that
this energy differs from the actual photon energy because the energy
levels are shifted by radiative corrections: the effect of these shifts
will be accounted for perturbatively below. The pole term is
\begin{equation}
\Sigma_{\rm pole} = \sum_b g_{vbbv}(\Delta E).
\end{equation}
In calculations of the Lamb shift the real part of this is taken,
but here we are concerned with the imaginary part, which gives the
lowest-order decay rate
\begin{equation}
\Gamma_0 = -2 \sum_b \bar{g}_{vbbv}(\Delta E),
\end{equation}
which can be written as a partial wave expansion
\begin{eqnarray}
\Gamma_0 = -8\pi\alpha \Delta E && \!\!\!\sum_{blm}
\int\! d^3 x\, j_l(\Delta E x) Y_{lm}(\Omega_x)\,
\bar{\psi}_v(\vec x) \gamma_{\mu} \psi_b(\vec x)
\nonumber \\ && \times
\int\! d^3 y\, j_l(\Delta E y) Y^*_{lm}(\Omega_y)\,
\bar{\psi}_b(\vec y) \gamma^{\mu} \psi_v(\vec y).
\label{basic0}
\end{eqnarray}
Because Feynman gauge has been used for the self-energy calculation,
this form of the decay rate is different from derivations
which use the properties of the actual transverse photons that are
emitted, but the result is the same because of gauge invariance.
In Table~\ref{tab1}, we present the lowest-order rates for the states
$2s_{1/2}$, $2p_{1/2}$, and $2p_{3/2}$ to decay by one-photon emission
to the ground state for $Z = 5, 10, \ldots, 100$. We now turn to the
radiative corrections to these decay rates, which we define in terms
of a function $R(Z\alpha)$,
\begin{equation}
\Gamma = \Gamma_0\, \Big[ 1 + {\alpha \over \pi}\, R(Z\alpha) \Big].
\end{equation}
Before presenting the exact calculation of $R(Z\alpha)$, we calculate
the leading contribution of order $(Z\alpha)^2$ for the $p$ states using
an effective field theory approach. We do not treat the more complicated
$s$ state correction, as the M1 decay is highly suppressed at low $Z$.

\section{Perturbative Approach}

When $Z$ is small, an expansion in $Z\alpha$ converges rapidly.
We present here the calculation of $R(Z\alpha)$ to leading order
$(Z\alpha)^2$, which will serve as a check of the nonperturbative
treatment presented in the next section. The radiative correction
to the decay rate is obtained from the nonrelativistic form of
quantum electrodynamics (QED) supplemented by one-loop corrections
to electron form factors $F_1$, $F_2$ and the vacuum polarization.
In the lowest order, the decay rate of the $2P$ state in
hydrogen-like atoms is
\begin{equation}
\Gamma_0 = \frac{4}{9}\, \alpha E^3
\left| \langle 1S\, | \vec{r}\, | 2\vec{P}\, \rangle \right|^2,
\label{basic1}
\end{equation}
where $E$ is the nonrelativistic limit of $\Delta E$ defined in
the previous section,
\begin{equation}
E = E(2P) - E(1S) = {3m(Z\alpha)^2 \over 8},
\end{equation}
with the nonrelativistic wave functions
\begin{eqnarray}
\phi_{1S} &=& \frac{1}{\sqrt{4\pi}}\,
(mZ\alpha)^{3/2}\, 2e^{-mZ\alpha r}, \\
\vec\phi_{2P} &=& \frac{1}{\sqrt{4\pi}}\, \frac{1}{2\sqrt{6}}\,
(mZ\alpha)^{3/2}\, e^{-mZ\alpha r/2}\, (mZ\alpha \vec r).
\end{eqnarray}
Natural units in which $\hbar = c = 1$ are used here.
Note that $\vec\phi_{2P}$ is normalized here in a nonstandard way,
namely $\int d^3 r\,\vec\phi_{2P}\cdot \vec\phi_{2P} = 1$.
Using the nonrelativistic matrix element
\begin{equation}
d \equiv \langle 1S\, | \vec{r}\, | 2\vec{P}\, \rangle
= \frac{1}{\sqrt{6}}\, \frac{256}{81}\, {1 \over mZ\alpha},
\end{equation}
Eq.~(\ref{basic1}) gives the well-known decay rate
\begin{equation}
\Gamma_0 = {2^8 \over 3^8}\, m\alpha(Z\alpha)^4.
\end{equation}
The radiative corrections to this can be expressed as
\begin{equation}
{\alpha \over \pi}\, R \equiv \frac{\delta \Gamma}{\Gamma_0}
= 3\,\frac{\delta E}{E} + 2\,\frac{\delta d}{d}\, .
\label{eq:18}
\end{equation}
When QED effects can be treated as local potentials, the calculation
of radiative corrections is relatively simple. We illustrate this with
the correction due to the presence of vacuum polarization, which
is given in the nonrelativistic limit by a local interaction potential
\begin{equation}
\delta V = -{4Z\alpha^2 \over 15m^2}\, \delta^3(\vec r).
\label{eq:19}
\end{equation}
Corrections to the energy and wave function of the $2P$ state from
$\delta V$ do not contribute to $R(Z\alpha)$ at the order of interest, 
but the potential shifts the $1S$ energy by
\begin{equation}
\delta E(1S) = -{4m\alpha(Z\alpha)^4 \over 15\pi},
\end{equation}
which gives a contribution to $R(Z\alpha)$ of $32/15\, (Z\alpha)^2$.
In addition the potential shifts the $1S$ wave function by
\begin{eqnarray}
\delta\phi_{1S} &\equiv& \bigg\langle r\,
\bigg| \frac{1}{(E-H)'}\, \delta V\, \bigg| 1S \bigg\rangle
\nonumber \\ &=&
{8\alpha(Z\alpha^2) \over 15\pi}\, e^{-mZ\alpha r}
\bigg[ {5 \over 2} - \gamma_E - mZ\alpha r +
{1 \over 2mZ\alpha r} - \ln(2mZ\alpha r) \bigg]
{1 \over \sqrt{4\pi}} (mZ\alpha)^{3/2},~~
\end{eqnarray}
which leads to a total contribution from vacuum polarization of
\begin{eqnarray}
R_{\rm VP}(Z\alpha) &=&
(Z\alpha)^2\, \biggl\{ {32 \over 15} + \biggl[ -{8 \over 15}
\ln{4 \over 3} - {131 \over 90} \biggr] \biggr\}
\nonumber \\ &=&
(Z\alpha)^2\, \biggl[ {61 \over 90} - {8 \over 15}
\ln{4 \over 3} \biggr]
\nonumber \\ &=&
(Z\alpha)^2\; \Bigl[\, 0.524\;347\, \Bigr].
\label{eq:pertvp}
\end{eqnarray}
We note the strong cancellation between the effect of the energy shift,
which is automatically accounted for when experimental energies are used,
and the perturbed orbital, which shows care needs to be taken when that
approach is taken. We now turn to the more complex self-energy correction.
The effect on the energy shift is well-known, coming from the self-energy
part of the Lamb shift of the $1S$ and $2P_{1/2}$ states of
\begin{eqnarray}
\delta E(1S) &=& \frac{m\alpha}{\pi}\, (Z\alpha)^4\,
\biggl[ \frac{10}{9} + \frac{4}{3} \ln(Z\alpha)^{-2} -
\frac{4}{3} \ln k_0(1S) \biggr],
\label{eq:23} \\
\delta E(2P) &=& \frac{m\alpha}{\pi}\, (Z\alpha)^4\, \frac{1}{8}\,
\biggl[ -\frac{1}{6} - \frac{4}{3} \ln k_0(2P) \biggr],
\label{eq:24}
\end{eqnarray}
where
\begin{eqnarray}
\ln k_0(1S) &=& 2.984\;128\;556\, , \\
\ln k_0(2P) &=& -0.030\;016\;709\, .
\end{eqnarray}
This energy shift contributes to the decay rate in accordance with
Eq.~(\ref{eq:18}). However, the radiative corrections to the dipole
matrix element are more difficult to obtain. We split this correction
into three parts, $\delta d = \delta d_L + \delta d_H + \delta d_K$, 
where $\delta d_L$ comes from low-energy photons, $\delta d_H$ 
is the high-energy correction to the wave function, and $\delta d_K$ is the 
correction to the dipole operator. 
Using nonrelativistic QED one derives the following expression for
$\delta d_L$
\begin{eqnarray}
\delta d_L &=& \frac{2\alpha}{3\pi m^2} \int_0^\epsilon\!
\omega\, d\omega\, \Re[f(\omega)]\, ,
\label{eq:27}
\\
f(\omega) &=&
\biggl\langle S \, \biggl|\, p_i\, \frac{1}{H - E_S + \omega}\, r_j\,
\frac{1}{H - E_P + \omega}\, p_i\, \biggr| P_j \biggr\rangle
\nonumber \\ && + \,
\biggl\langle S \, \biggl|\, r_j\, \frac{1}{(H - E_P)'}\, p_i\,
\frac{1}{H - E_P + \omega}\, p_i\, \biggr| P_j \biggr\rangle
\nonumber \\ && + \,
\biggl\langle S \, \biggl|\, p_i\, \frac{1}{H - E_S + \omega}\, p_i\,
\frac{1}{(H - E_S)'}\, r_j\, \biggr| P_j \biggr\rangle
\nonumber \\ && - \,
\frac{d}{2}\, \biggl\langle S \, \biggl|\, p_i\,
\frac{1}{(H - E_S + \omega)^2}\, p_i\, \biggr| S \biggr\rangle
\nonumber \\ && - \,
\frac{d}{2}\, \biggl\langle P_j\, \biggl|\, p_i\,
\frac{1}{(H - E_P + \omega)^2}\, p_i\, \biggr |P_j \biggr\rangle\, ,
\end{eqnarray}
where $\epsilon$ is assumed to be asymptotically large
and $i$, $j$ are vector coordinate indices.
All matrix elements of $f(\omega)$ are calculated numerically using a
finite difference representation of the nonrelativistic Hamiltonian.
The integration with respect to $\omega$ requires special treatment
regarding linear and logarithmic in $\epsilon$ terms. The large
$\omega$ asymptotics is
\begin{equation}
f(\omega) \approx \frac{A}{\omega} + \frac{B}{\omega^2} +
\frac{C}{\omega^{5/2}} + \ldots
\end{equation}
where
\begin{eqnarray}
A &=& \biggl\langle S \, \biggl|\, p^2\,
\frac{1}{(H - E_S)'}\, r_i\, \biggr| P_i \biggr\rangle\, +\,
\biggr\langle S \, \biggl|\, r_i\,
\frac{1}{(H - E_P)'}\, p^2\, \biggr| P_i \biggr\rangle = -2d
\\
B &=& -\frac{1}{2}\,
\biggl\langle S \, \biggl|\, 4\pi Z\alpha\,\delta^3(r)\,
\frac{1}{(H - E_S)'}\, r_i\, \biggr| P_i \biggr\rangle =
d\,(Z\alpha)^2\, \biggl( \frac{131}{24} + 2\ln\frac{4}{3} \biggr),
\\
C &=& 2\sqrt{2}B\, (Z\alpha).
\end{eqnarray}
The numerical integration in Eq.~(\ref{eq:27}), along a contour
which omits poles from above or below, leads to the result
\begin{equation}
\delta d_L = \frac{2\alpha}{3\pi}\, \biggl\{ A \epsilon +
B \ln\biggl[\frac{2\epsilon}{(Z\alpha)^2}\biggr] -
d\, (Z\alpha)^2\, 17.759\;359 \biggr\}\, .
\end{equation}
The term linear in $\epsilon$ is dropped, and the logarithmic
term is canceled by the contribution $\delta d_H$ coming from
large photon momenta. This latter contribution can be expressed in terms of an
interaction potential $\delta V$ obtained from the one-loop
electron form factors $F_1$ and $F_2$,
\begin{equation}
\delta V = Z\alpha^2 \biggl[ \frac{10}{9} -
\frac{4}{3} \ln(2\epsilon) \biggr] \delta^3(r) +
\frac{Z\alpha^2}{2\pi}\, \frac{\vec{L} \cdot \vec{S}}{r^3}\, .
\end{equation}
It contributes to the energy shift in a way that has already been accounted
for in Eqs.~(\ref{eq:23}, \ref{eq:24}), but also gives corrections
to the wave functions, and thus to the transition dipole moment
\begin{eqnarray}
\delta d_H &=& Z\alpha^2 \biggl[ \frac{10}{9} -
\frac{4}{3} \ln(2\epsilon) \biggr] \biggl\langle S \,
\biggl| \delta^3(r)\, \frac{1}{(E_S - H)'}\, \vec{r}\, \biggr|
\vec{P} \biggr\rangle
\nonumber \\ && + \,
\frac{Z\alpha^2}{2\pi}\, \biggl\langle S\, \biggl|\, \vec{r}\,
\frac{1}{(E_P - H)'}\, \frac{\vec{L} \cdot \vec{S}}{r^3}\, \biggr|
\vec{P} \biggr\rangle.
\end{eqnarray}
There is one more spin-dependent term which recently was 
discussed in Ref. \cite{forbidden}. It arises from the anomalous magnetic 
moment $\kappa$ correction to the dipole transition operator
\begin{equation}
i\,\omega\,\vec r -\frac{\kappa}{4\,m}\,
\vec k^2\,\vec r\times\vec\sigma.
\end{equation}
Its matrix element between $S$ and $P_{1/2}$ states 
leads to a correction $\delta d_K$
\begin{equation}
\delta d_K = -d\,\frac{E\,\kappa}{4\,m}\,2  = 
-d\,\frac{3\,\alpha}{32\,\pi}\,(Z\,\alpha)^2.
\end{equation}
With the help of Eq.~(\ref{eq:18}), the sum
$\delta d_L + \delta d_H + \delta d_K$, 
together with energy shift contributions
from Eqs.~(\ref{eq:23}, \ref{eq:24}), leads finally to the result for
the radiative correction to the decay rate of the $2P_{1/2}$ state
\begin{equation}
R_{\rm SE}^{2p_{1/2}}(Z\alpha) = (Z\alpha)^2 \biggl\{
\biggl[ {8 \over 3}\, \ln{4 \over 3} - {61 \over 18} \biggr]\!
\ln(Z\alpha)^{-2} + 6.051~68 \biggr\} .
\label{eq:pert1/2}
\end{equation}
A similar calculation for the $2P_{3/2}$ state yields
\begin{equation}
R_{\rm SE}^{2p_{3/2}}(Z\alpha) = (Z\alpha)^2 \biggl\{
\biggl[ {8 \over 3}\, \ln{4 \over 3} - {61 \over 18} \biggr]\!
\ln(Z\alpha)^{-2} + 5.984~36 \biggr\} .
\label{eq:pert3/2}
\end{equation}
The coefficient of the logarithmic term is in agreement with
\cite{Ivanov}.

\section{Two-loop formalism}

\begin{figure}[p]
\centerline{\includegraphics[scale=1.0]{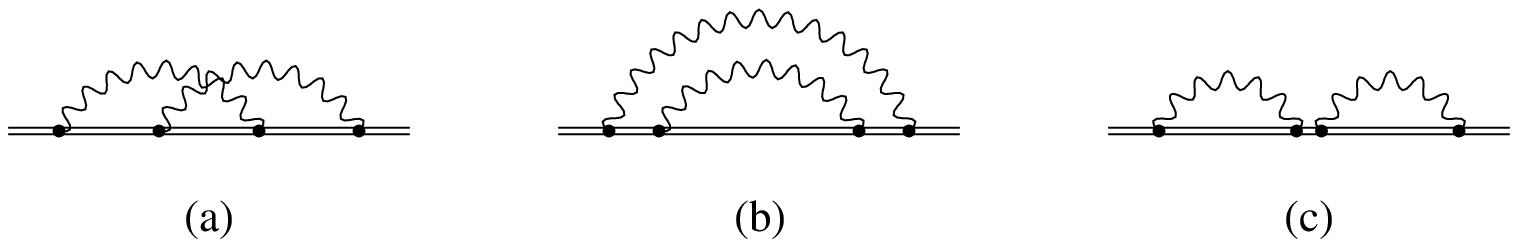}}
\caption{\label{fig1}
Two-loop Lamb shift diagrams.}
\end{figure}

Following the approach given above to calculate radiative corrections
to decay rates, we consider the imaginary part of the two-loop Lamb shift.
We begin by considering the three self-energy diagrams of Fig.~\ref{fig1},
leaving vacuum polarization for later. Expressions for the diagrams,
which we refer to as overlap, nested, and reducible following the
notation of Fox and Yennie \cite{Fox-Yennie}, were derived by
Mills and Kroll \cite{Kroll-Mills}, and we now treat them in order.

\subsection{Overlap diagram}

The overlap diagram, Fig.~\ref{fig1}a, is given by
\begin{eqnarray}
\Sigma^{4O} &=& -e^4 \!\int\! d^3 x\, d^3 y\, d^3 z\, d^3 w\!
\int\! {d^n k \over (2 \pi)^n}\, {d^n l \over (2 \pi)^n}\,
{e^{i \vec k \cdot (\vec x - \vec z)} \over k^2 + i\delta}\,
{e^{i \vec l \cdot (\vec y - \vec w)} \over l^2 + i\delta}\,
\bar{\psi}_v(\vec x) \gamma^{\mu}
\nonumber \\ && \times \,
S_F(\vec x, \vec y; \epsilon_v - k_0) \gamma^{\nu}
S_F(\vec y, \vec z; \epsilon_v - k_0 - l_0) \gamma_{\mu}
S_F(\vec z, \vec w; \epsilon_v - l_0) \gamma_{\nu} \psi_v(\vec w).
\end{eqnarray}
As with the one-loop case, we introduce spectral representations for the
electron propagators and carry out the $d^3 k$ and $d^3 l$ integrations
to get
\begin{equation}
\Sigma^{4O} = - \sum_{mnr} \int {dk_0 \over 2\pi}
\int {dl_0 \over 2 \pi} \, {g_{vnmr}(k_0)\, g_{mrnv}(l_0) \over
(\epsilon_v - k_0 - \epsilon_m) (\epsilon_v - k_0 -l_0 - \epsilon_n)
(\epsilon_v - l_0 - \epsilon_r)},
\end{equation}
where in this section we leave the factor $(1 - i\delta)$
multiplying energies in the spectral representation of the
electron propagator understood. We now consider Wick rotating both
$k_0 \rightarrow i \omega_k$ and $l_0 \rightarrow i \omega_l$.
If no poles are passed, this again leads to a purely real expression.
To get an imaginary part, at least one of the three denominators must
involve encircling a pole. The middle denominator can have a pole when
$n=a$ and $k_0 + l_0 = \Delta E$, but this corresponds to two-photon
decay which we do not treat here. We then need consider only the
two cases when either $m=a$ and $k_0 = \Delta E$ or $r=a$ and
$l_0 = \Delta E$, which gives the expressions
\begin{eqnarray}
\Sigma^{4O}_L &=& -4\pi i\alpha^2 \sum_b
\int\! d^3 x\, d^3 y\, d^3 z\, d^3 w\,
{e^{i \Delta E |\vec x - \vec z|} \over |\vec x - \vec z|}
\!\int\! {d^n l \over (2\pi)^n} \,
{e^{i \vec l \cdot (\vec y - \vec w)} \over l^2 + i\delta}
\nonumber \\ && \times \,
\bar{\psi}_v(\vec x) \gamma^{\mu} \psi_b(\vec x) \,
\bar{\psi}_b(\vec y) \gamma^{\nu}
S_F(\vec y, \vec z; \epsilon_a - l_0) \gamma_{\mu}
S_F(\vec z, \vec w; \epsilon_v - l_0) \gamma_{\nu} \psi_v(\vec w)
\end{eqnarray}
and
\begin{eqnarray}
\Sigma^{4O}_R &=& -4\pi i\alpha^2 \sum_b
\int\! d^3 x\, d^3 y\, d^3 z\, d^3 w\,
{e^{i \Delta E |\vec y - \vec w|} \over |\vec y - \vec w|}
\!\int\! {d^n k \over (2 \pi)^n} \,
{e^{i \vec k \cdot (\vec x - \vec z)} \over k^2 + i\delta}
\nonumber \\ && \times \,
\bar{\psi}_v(\vec x) \gamma^{\mu}
S_F(\vec x, \vec y; \epsilon_v - k_0) \gamma^{\nu}
S_F(\vec y, \vec z; \epsilon_a - k_0) \gamma_{\mu} \psi_b(\vec z) \,
\bar{\psi}_b(\vec w) \gamma_{\nu} \psi_v(\vec w),
\end{eqnarray}
where we have ``undone'' the spectral representations of the electron
propagator and kept either the $d^3 k$ or $d^3 l$ integration.

We note at this point that these expressions are almost identical to
expressions that arise in the treatment of screening corrections to the
 self-energy in lithiumlike ions (Eqs.~(25, 27) in Ref.~\cite{libi}),
with the only difference being an overall minus sign and the fact that
we are interested in the imaginary part here, while the real part was
calculated in \cite{libi}. We were able then, with only slight
modifications, to use code developed for the screening corrections
in lithiumlike ions for the present calculation. Replacing
\begin{equation}
  e^{i\Delta E |\vec y - \vec w|} \rightarrow
i\sin(\Delta E |\vec y - \vec w|)
\end{equation}
and using the equality of the two terms gives the net result for the
decay rate contribution from the overlap diagram we call $\Gamma_V$,
\begin{eqnarray}
\Gamma_V &=& -16\pi\alpha^2 \Im \sum_b
\int\! d^3 x\, d^3 y\, d^3 z\, d^3 w\,
{\sin(\Delta E |\vec x - \vec z|) \over |\vec x - \vec z|} \,
\bar{\psi}_v(\vec x) \gamma^{\mu} \psi_b(\vec x)
\nonumber \\ && \times
\int\! {d^n k \over (2\pi)^n} \,
{e^{i \vec k \cdot (\vec y - \vec w)} \over k^2 + i\delta} \,
\bar{\psi}_b(\vec y) \gamma^{\nu}
S_F(\vec y, \vec z; \epsilon_a - k_0) \gamma_{\mu}
S_F(\vec z, \vec w; \epsilon_v - k_0) \gamma_{\nu} \psi_v(\vec w).
\label{overlap}
\end{eqnarray}

\subsection{Nested diagram}

The nested diagram, Fig.~\ref{fig1}b, is given by
\begin{eqnarray}
\Sigma^{4N} &=& -e^4 \!\int\! d^3 x\, d^3 y\, d^3 z\, d^3 w\!
\int\! {d^n k \over (2\pi)^n}\, {d^n l \over (2\pi)^n}\,
{e^{i \vec k \cdot (\vec x - \vec w)} \over k^2 + i\delta}\,
{e^{i \vec l \cdot (\vec y - \vec z)} \over l^2 + i\delta}\,
\bar{\psi}_v(\vec x) \gamma^{\mu}
\nonumber \\ && \times \,
S_F(\vec x, \vec y; \epsilon_v - k_0) \gamma^{\nu}
S_F(\vec y, \vec z; \epsilon_v - k_0 - l_0) \gamma_{\nu}
S_F(\vec z, \vec w; \epsilon_v - k_0) \gamma_{\mu} \psi_v(\vec w).
\end{eqnarray}
This leads to the expression
\begin{equation}
\Sigma^{4N} = - \sum_{mnr}
\int {dk_0 \over 2\pi} \int {dl_0 \over 2 \pi}\,
{g_{vrmv}(k_0)\, g_{mnnr}(l_0) \over (\epsilon_v - k_0 - \epsilon_m)
(\epsilon_v - k_0 -l_0 - \epsilon_n) (\epsilon_v - k_0 - \epsilon_r)}.
\end{equation}
We again consider Wick rotating both $k_0 \rightarrow i\omega_k$ and
$l_0 \rightarrow i\omega_l$, which gives a real result if no poles are
passed. To get an imaginary part, at least one of the three denominators
must encircles a pole, and once again, we omit poles arising from the
middle denominator, which correspond to two-photon decay. We therefore
need to consider only the $k_0$ Wick rotation, which has poles when
$k_0 = \Delta E$ and either $m=a$ or $r=a$. However, if both $m$ and $r$
are the ground state, a double pole is encountered. If the double pole
is excluded, two terms result,
\begin{equation}
\Sigma^{4N}_L = \sum_{br}^{r \neq a} {g_{vrbv}(\Delta E)
\Sigma_{br}(\epsilon_a) \over \epsilon_a - \epsilon_r}\, ,
\end{equation}
and
\begin{equation}
\Sigma^{4N}_R = \sum_{bm}^{m \neq a} {g_{vbmv}(\Delta E)
\Sigma_{mb}(\epsilon_a) \over \epsilon_a - \epsilon_m}\, .
\end{equation}
These terms can be written in terms of perturbed orbitals.
Specifically, if we define
\begin{equation}
\tilde{\psi}_a(\vec z) \equiv \sum_{br}^{r \neq a} \psi_r(\vec z)\,
{g_{vrbv}(\Delta E) \over \epsilon_a - \epsilon_r}\, \delta_{m_b m_r},
\end{equation}
and
\begin{equation}
\tilde{\bar{\psi}}_a(\vec z) \equiv \sum_{bm}^{m \neq a}
\bar{\psi}_m(\vec z)\, {g_{vbmv}(\Delta E) \over
\epsilon_a - \epsilon_m}\, \delta_{m_b m_m},
\end{equation}
then
\begin{equation}
\Sigma^{4N}_L + \Sigma^{4N}_R =
\Sigma_{a \tilde{a}}(\epsilon_a) + \Sigma_{\tilde{a} a}(\epsilon_a).
\label{qsl2}
\end{equation}
Because the ground state self-energy is purely real, the only contribution
to the decay rate comes from the imaginary part of these perturbed orbital
terms.

To treat the double pole, we set $m=b$ and $r=c$, and find
\begin{equation}
\Sigma^{4N}_D = i \sum_{bc} \int {dk_0 \over 2\pi}\, {g_{vcbv}(k_0)
\Sigma_{bc}(\epsilon_v - k_0) \over (\Delta E - k_0 + i\delta)^2}\, .
\end{equation}
Applying Cauchy's theorem and using the fact that the self-energy is
diagonal in magnetic quantum numbers gives two derivative terms,
\begin{equation}
\Sigma^{4N}_D =
- \sum_b g'_{vbbv}(\Delta E) \Sigma_{aa}(\epsilon_a)
+ \sum_b g_{vbbv}(\Delta E) {\Sigma'}_{\!\!aa}(\epsilon_a).
\label{4Nder}
\end{equation}

\subsection{One-Particle Reducible Diagram}

The final contribution to the two-loop self-energy comes from
Fig.~\ref{fig1}c, which breaks into two parts, a perturbed orbital term
\begin{equation}
\Sigma_{PO} = \sum_{m\neq v} {\Sigma_{vm}(\epsilon_v)
\Sigma_{mv}(\epsilon_v) \over \epsilon_v - \epsilon_m} ,
\end{equation}
and a derivative term
\begin{equation}
\Sigma_D = \Sigma_{vv}(\epsilon_v) \,
{\partial \Sigma_{vv}(E) \over \partial E} \,\Bigg|_{E = \epsilon_v}.
\end{equation}
The perturbed orbital term will have an imaginary part only if at least
one pole term is present, as our analysis of the complex nature of
$\Sigma$ did not depend on the external wavefunctions, so long as they
are real. This then leads to an imaginary contribution to the energy of
\begin{equation}
\Sigma_{PO}(a) = i [ \Sigma_{v \tilde{v}}(\epsilon_v)
                   + \Sigma_{\tilde{v} v}(\epsilon_v) ],
\label{qsl1}
\end{equation}
where
\begin{equation}
\tilde{\psi}_v(\vec z) \equiv
\sum_{br}^{r \neq v} \psi_r(\vec z)\,
{\bar{g}_{vrbv}(\Delta E) \over \epsilon_v - \epsilon_r}\, ,
\end{equation}
and
\begin{equation}
\tilde{\bar{\psi}}_v(\vec z) \equiv
\sum_{bm}^{m \neq v} \bar{\psi}_m(\vec z)\,
{\bar{g}_{vbmv}(\Delta E) \over \epsilon_v - \epsilon_m}\, .
\end{equation}

The derivative term will lead to an imaginary part of the energy in two
ways: in the first, we take the imaginary part of the first self-energy,
which is of course associated with the lowest-order decay rate, and
multiply it by the real part of the derivative of the valence self-energy.
We combine this term with the first term of Eq.~(\ref{4Nder}) to get the
``derivative A'' term,
\begin{equation}
\Gamma_{dera} = \Gamma_0\, [{\Sigma'}_{\!\!aa}(\epsilon_a) +
                        \Re {\Sigma'}_{\!\!vv}(\epsilon_v)].
\label{dera}
\end{equation}
The second contribution is when the real part of the first self-energy
multiplies the imaginary part of the derivative of the self-energy,
which can be combined with the second part of Eq.~(\ref{4Nder}) to give
\begin{equation}
\Gamma_{derv} = \Gamma_0'\, [\,\Re \Sigma_{vv}(\epsilon_v) -
                                   \Sigma_{aa}(\epsilon_a)].
\label{derv}
\end{equation}
In our numerical analysis, we simply evaluate $\Gamma_0'$ as one
object. However, as can be seen by referring to Eq.~(\ref{basic0}),
a multiplicative factor $\Delta E$ is present in the formula for
$\Gamma_0$. If the derivative acts on this term, a contribution of
$\Gamma_0/\Delta E$ would be present, as is the case in the formulas
given by Barbieri and Sucher \cite{Barbieri-Sucher}.

\section{Vacuum polarization}

While the exact treatment of vacuum polarization is somewhat complicated,
to order $\alpha (Z\alpha)^2$ one needs to consider only the analog of
the $1s$ perturbed orbital. This is to be contrasted with the effective
field theory discussion, in which both that perturbed orbital and an energy
shift needed to be considered. While the effect of the energy shift is
present in the exact calculation, which enters through the analog of the
derivative A term, it is a peculiarity of the Feynman gauge that the
low-$Z$ behavior of $\Gamma_0'$ is of order $(Z\alpha)^4$: this arises through
a cancellation between timelike and spacelike terms, which separately behave
as $(Z\alpha)^2$. Replacing the self-energy with the Uehling potential in
the $1s$ perturbed orbital gives numerical results that are consistent with
Eq.~(\ref{eq:pertvp}).

\section{Rearrangement for Numerical Evaluation}

In this section we perform further manipulations on the basic expressions
for the two-loop self-energy that will allow an exact numerical evaluation.
Beginning with the overlap term, we note that it is ultraviolet divergent.
We deal with that divergence by considering Eq.~(\ref{overlap}) with the
bound state propagators replaced with free propagators, which leads to an
expression we denote $\Gamma_{V1}$,
\begin{eqnarray}
\Gamma_{V1} &=& -{2\alpha^2 \over \pi^2}\,
\Im \!\int\! d^3 x \!\int\! d^3 z\,
{\sin(\Delta E |\vec x - \vec z|) \over |\vec x - \vec z|}
\sum_b \bar{\psi}_v(\vec x) \gamma^{\mu} \psi_b(\vec x)
\!\int\! d^3 p_2 \int\! d^3 p_1
\nonumber \\ && \times \,
e^{i \vec z \cdot (\vec p_1 - \vec p_2)}
\!\int\! {d^n k \over (2\pi)^n}\, {1 \over k^2 + i\delta}\,
\bar{\psi}_b(\vec p_2) \gamma_{\nu}
{1 \over \not\!p_2 - \not\!k - m} \gamma_{\mu}
{1 \over \not\!p_1 - \not\!k - m} \gamma^{\nu} \psi_v(\vec p_1),
\end{eqnarray}
where $p_2 = (\epsilon_a,\ \vec p_2)$ and $p_1 = (\epsilon_v,\ \vec p_1)$.
If we define $\vec q = \vec p_2 - \vec p_1$ and
\begin{equation}
J^{\mu}_{vb}(\vec q) = \!\int\! d^3 x\, \bar{\psi}_v(\vec x)
\gamma^{\mu} \psi_b(\vec x) e^{-i \vec x \cdot \vec q},
\end{equation}
this can be rewritten as
\begin{eqnarray}
\Gamma_{V1} &=& - {4\alpha^2 \over \Delta E}\,
\Im \sum_b \int\! d^3 p_2\, d^3 p_1 J^{\mu}_{vb}(\vec q)\,
\delta(|\vec q\,| - \Delta E)
\!\int\! {d^n k \over (2\pi)^n}\, {1 \over k^2 + i \delta}
\nonumber \\ && \times \,
\bar{\psi}_b(\vec p_2) \gamma^{\nu}
{1 \over \not\!p_2 - \not\!k - m} \gamma_{\mu}
{1 \over \not\!p_1 - \not\!k - m} \gamma_{\nu} \psi_v(\vec p_1).
\end{eqnarray}
Standard Feynman diagram techniques can now be used to write this as
\begin{eqnarray}
\Gamma_{V1} &=& {\alpha C (1 - \epsilon) \over \pi \epsilon}\, \Gamma_0
\nonumber \\ && + \,
{\alpha^2 \over 2\pi^2 \Delta E} \sum_b \int\! d^3 p_2\, d^3 p_1
J^{\mu}_{vb}(\vec q)\, \delta(|\vec q\,| - \Delta E)
\!\int_0^1\! \rho d \rho \!\int_0^1\! dx\,
N_{0\mu} \ln{\Delta \over m^2}
\nonumber \\ && + \,
{\alpha^2 \over 4\pi^2 \Delta E} \sum_b \int\! d^3 p_2\, d^3 p_1
J^{\mu}_{vb}(\vec q)\, \delta(|\vec q\,| - \Delta E)
\!\int_0^1\! \rho d \rho \!\int_0^1\! dx\, {N_{\mu} \over \Delta},
\end{eqnarray}
where
\begin{eqnarray}
C &=& (4\pi)^{\epsilon/2} \Gamma(1+\epsilon/2),
\nonumber \\
Q &=& \rho [x p_1 + (1-x) p_2],
\nonumber \\
\Delta &=& \rho x (m^2 - p_1^2) + \rho(1-x) (m^2 - p_2^2) + Q^2,
\nonumber \\
N_{0\mu} &=& \bar{\psi}_b(\vec p_2) \gamma_{\mu} \psi_v(\vec p_1),
\nonumber
\end{eqnarray}
and
\begin{equation}
N_{\mu} = \bar{\psi}_b(\vec p_2) \gamma_{\nu}
(\not\!p_2 - \not\!Q + m) \gamma_{\mu}
(\not\!p_1 - \not\!Q + m) \gamma^{\nu} \psi_v(\vec p_1).
\end{equation}
We note that the momentum space form of the lowest-order decay rate is
\begin{equation}
\Gamma_0 = -{\alpha \over 2\pi \Delta E} \sum_b \int\! d^3 p_2\, d^3 p_1
J^{\mu}_{vb}(\vec q)\, \delta(|\vec q\,| - \Delta E)\, N_{0{\mu}}.
\end{equation}
The ultraviolet divergent term in $\Gamma_{V1}$ will be shown to cancel
with derivative terms below, and the finite remainder is tabulated as
$Q_{V1}$ in the second columns of Tables \ref{tab2} -- \ref{tab4},
where we adopt the convention
\begin{equation}
\Gamma_x = {\alpha \over \pi}\, \Gamma_0\, Q_x.
\end{equation}

We can now deal with an ultraviolet finite expression by evaluating
$\Gamma_{V} - \Gamma_{V1}$. To numerically evaluate the subtracted form,
we first carry out the Wick rotation $k_0 \rightarrow i\omega$. If this
passes no poles, it is straightforward to show that an expression we
refer to as $\Gamma_{V2}$ results,
\begin{eqnarray}
\Gamma_{V2} &=& {4\alpha^2 \over \pi}
\sum_b \int\! d^3 x\, d^3 y\, d^3 z\, d^3 w\,
{\sin(\Delta E |\vec x - \vec z|) \over |\vec x - \vec z|}\,
\Re \!\int_0^{\infty}\!\! d\omega\,
{e^{-\omega |\vec y - \vec w|} \over |\vec y - \vec w|}
\nonumber \\ && \times \,
\bar{\psi}_v(\vec x) \gamma_{\mu} \psi_a(\vec x)\,
\bar{\psi}_a(\vec y) \gamma_{\nu}
S_F(\vec y, \vec z; \epsilon_a - i\omega) \gamma^{\mu}
S_F(\vec z, \vec w; \epsilon_v - i\omega) \gamma^{\nu} \psi_v(\vec w),
\end{eqnarray}
where a subtraction of the same form with free electron propagators is
understood. A kind of infrared divergence called a reference state
singularity is present in the above, and is regulated by taking
$\epsilon_v \rightarrow \epsilon_v (1 - \delta)$ and
$\epsilon_a \rightarrow \epsilon_a (1 - \delta)$, where $\delta$ is
chosen here to be $10^{-6}$. $\Gamma_{V2}$ has a logarithmic dependence
on $\delta$ which cancels with derivative terms and this is one of the
checks used in the calculation. It is possible to combine the terms
together to manifest the cancellation, but we have found it simpler
to work with a small, finite value of $\delta$, checking of course that
the sum is unchanged when $\delta$ is varied. We tabulate $Q_{V2}$ in
the third columns of Tables \ref{tab2} -- \ref{tab4}.

Finally, the Wick rotation picks up pole terms. To treat these, it is
convenient to rewrite Eq.~(\ref{overlap}) as
\begin{equation}
\Gamma_V = 4\, \Im \!\int\! {dk_0 \over 2 \pi} \sum_{bmn}
{\bar{g}_{vmbn}(\Delta E)\, g_{bnmv}(k_0) \over
(\epsilon_a - k_0 - \epsilon_m) (\epsilon_v - k_0 - \epsilon_n)}\, .
\end{equation}
Because of the regularization procedure, the first term
in the denominator has no poles, but the second does when
$\epsilon_n < \epsilon_v$, which leads to the pole term
\begin{equation}
\Gamma_{QV3} = 4 \sum_{bmn}
{\bar{g}_{vmbn}(\Delta E)\, g_{bnmv}(\epsilon_v - \epsilon_n)
\over \epsilon_m + \epsilon_v - \epsilon_a - \epsilon_n} F_n\, ,
\end{equation}
where $F_n = 1$ for positive energy states with
$\epsilon_n < \epsilon_v$ and $F_n = 0$ otherwise.
The associated contribution $Q_{V3}$ is tabulated in
the fourth columns of Tables \ref{tab2} -- \ref{tab4}.

Evaluation of the derivative A terms of Eq.~(\ref{dera}) is similar
to that of $Q_{V2}$, as in both cases ultraviolet and reference state
singularities are present. The same procedures are used to deal with
this, namely a subtraction of a free-propagator term and use of the
$\delta$ regulator. The analog of $Q_{V3}$ is also present, although
in this case it involves a double pole. Since we have discussed the
evaluation of these derivative terms in some detail in a number
of other papers (see, e.g., \cite{libi,CS}),
here we simply combine the various finite effects into the terms
$Q_{SL2}$ and $Q_{SR2}$ in the tables, where $Q_{SL2}$ refers to
${\Sigma'}_{\!\!vv}$ and $Q_{SR2}$ to ${\Sigma'}_{\!\!aa}$.
An ultraviolet divergent term in the free-propagator terms cancels
the first term in $Q_{V1}$. The perturbed orbital terms are evaluated
using techniques for evaluation of the one-loop Lamb shift \cite{CJS},
with Eq.~(\ref{qsl1}) tabulated as $Q_{SL1}$ and Eq.~(\ref{qsl2}) as
$Q_{SR1}$. Finally, the derivative B term of Eq.~(\ref{derv}) is
tabulated as $Q_{derb}$.

\section{discussion}

The most numerically striking feature of the present calculation
is the very large degree of cancellation present in the 2s M1 decays,
which prohibits going below $Z=50$. In the lowest-order calculation,
while using Feynman gauge gives the correct answer, a large cancellation
between a timelike and spacelike contribution is present, leaving the
highly suppressed $(Z\alpha)^{10}$ result shown in Table \ref{tab1}.
This cancellation is lost in individual contributions to the radiative
correction, and is only restored in the sum. This strong cancellation
in fact served as a useful test of the formulas and numerical methods.
In the unlikely event that radiative corrections needed to be considered
for M1 decays in hydrogenic ions with lower $Z$, the calculation would
be better carried out in the Coulomb gauge.

\begin{figure}[p]
\centerline{\includegraphics[scale=0.8]{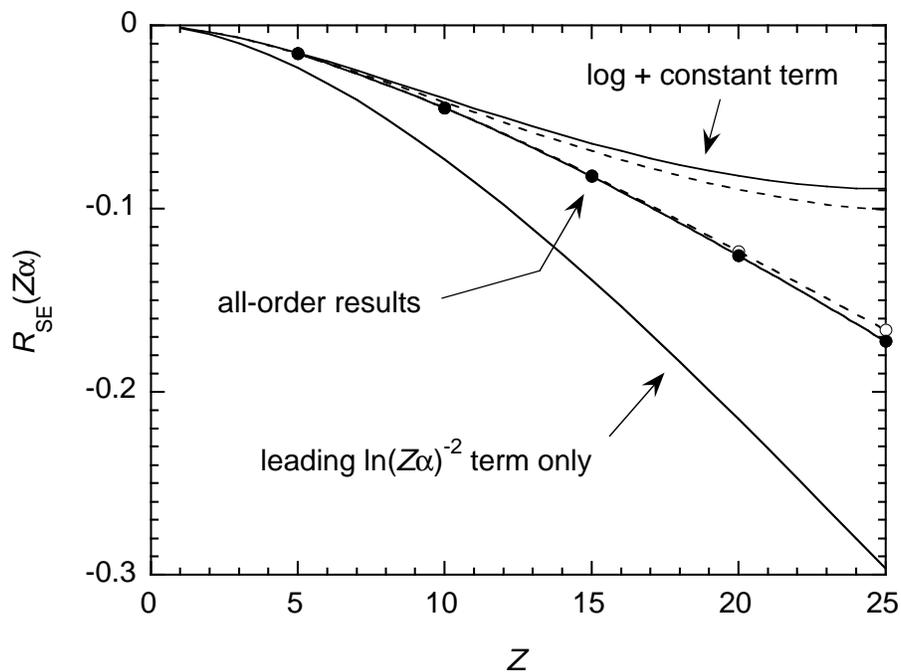}}
\caption{\label{fig2}
Comparisons between all-order and perturbative results of
$R_{\text{SE}}(Z\alpha)$. Solid and dashed lines are $2p_{1/2}$
and $2p_{3/2}$ results, respectively. Closed and open circles are
$2p_{1/2}$ and $2p_{3/2}$ all-order results, respectively.}
\end{figure}

Turning to the $2p$ E1 decay rates, we note that, while less severe
than for M1 decays, there is still considerable cancellation present
between the various contributions, particularly at low $Z$. This is
of course required by the fact that the $Z\alpha$ expansion series
for $R(Z\alpha)$ has no constant term, but instead starts in order
$(Z\alpha)^2$. Again, the high degree of cancellation between contributing
terms at low $Z$ serves as a check of our numerical calculations,
but in this case, we can also compare our low-$Z$ results with the
perturbation series. In Fig.~\ref{fig2}, all-order results of
$R_{\text{SE}}(Z\alpha)$ for $2p_{1/2}$ and $2p_{3/2}$ from Tables
\ref{tab3} and \ref{tab4} are compared with analytic results from
Eqs.~(\ref{eq:pert1/2}) and (\ref{eq:pert3/2}). It can be seen that
all-order results do converge to the analytic results at low $Z$.
In particular, the leading logarithmic term in the perturbation
series, which is the same for both $2p_{1/2}$ and $2p_{3/2}$,
is good for $Z=1$ and 2 only, while including the constant terms,
which leads to the splitting of fine structure results, extends the
validity of the perturbation series to $Z=7$ or 8. This is typical
of self-energy calculations where the $Z\alpha$ series is known to
converge very slowly except at very low $Z$ and nonperturbative
methods such as those shown here are needed for mid- to high-$Z$ ions.

In spite of the apparent agreement between the perturbation and all-order
results shown in Fig.~\ref{fig2}, it should be noted that there are
residual, unresolved discrepancies between them. By extending the
accuracy of our all-order calculations for $R_{\text{SE}}(Z\alpha)$ to
a level of $\pm$0.0002 for low-$Z$ ions, we were able to extract values
of 6.67 and 6.62 for the constant terms of the $2p_{1/2}$ and $2p_{3/2}$
states, respectively. While these results are uncertain to $\pm 0.20$
due to the high degree of numerical cancellation at low $Z$ when the
leading logarithmic term is subtracted out, they are nevertheless
different from the corresponding analytic values of 6.051~68 and
5.984~36 from Eqs.~(\ref{eq:pert1/2}) and (\ref{eq:pert3/2}).
Until this discrepancy is resolved, we would assign a 10\% error
to the constant terms, which should have negligible effect on
$R_{\text{SE}}(Z\alpha)$ anyway.

While the decay rate corrections here are of intrinsic interest,
the purpose of the present calculation is actually to serve as the
first step in the calculation of PNC corrections. There is interest
in the parity forbidden transition $6s_{1/2} \rightarrow 7s_{1/2}$
in cesium, which serves as one test of the electroweak part of the
standard model of particle physics. A very large radiative correction
has been found for this case \cite{Kuchiev}, \cite{Milstein}, but the only calculation
using exact propagators that has so far been carried out was for the
$2s_{1/2}-2p_{1/2}$ matrix element for hydrogenlike ions \cite{PNC}.
This calculation had the advantage of being gauge invariant
because of the degeneracy of the Dirac energies of the two states.
The calculation carried out in this paper is also gauge invariant
despite the differing energies of the $n=2$ states and the ground
state, and is generalizable to the PNC case. To carry out this
generalization, the extra perturbation of the effect of $Z$ boson
exchange between the nucleus and electron must be added.

An additional feature that must be dealt with for extending the
present calculation to neutral cesium is the complication of dealing
with a many-electron system. Because we use numerical Green's functions,
there is no difficulty in using a modified Furry representation of
QED in which the Coulomb potential is replaced with a model potential
that incorporates the dominant effect of electron screening. However,
the effect of the filled xenon-like core will have to be taken into
account, which will lead to extra diagrams. We are at present setting
up calculations of radiative corrections to allowed transitions in the
alkalis, in which these issues will arise, with the next step being
the inclusion of the effect of $Z$ exchange.

The challenges to experiment in testing the calculations presented here
are considerable. The largest radiative corrections found here are those
to M1 decays at high $Z$, with the correction at $Z=100$ being 1.2\%.
For E1 decays, even the largest case, $2p_{1/2}$ at $Z=100$, has a
radiative correction of only 0.2\%. Rather than a direct measurement,
experiments involving interference, such as the one discussed in
Ref.~\cite{Dunford}, may be more promising.

There is a radiative correction, even though very small at low $Z$,
that is of particular interest. It involves the E1 decay of the $2p_{1/2}$
state in hydrogen. One approach to the determination of the Lamb shift as
described in Ref.~\cite{Sok} involves the measurement of the decay rate
of the $2p_{1/2}$ state in hydrogen to very high accuracy. To interpret
the experiment, Ref.~\cite{Sok} used the following formula for the
radiative correction
\begin{equation}
R_{2p_{1/2}}(Z\alpha) \Big|_{\text{Ref.~\cite{Sok}}}
= {32 \over 3}\,
(Z\alpha)^2\, \bigg[ -\ln(Z\alpha)^{-2} - {1 \over 8} \ln k_0(2P) +
\ln k_0(1S) - {1 \over 64} - {19 \over 30}\, \bigg],
\end{equation}
which can be shown to be equivalent to the first term of Eq.~(\ref{eq:18}).
However, as discussed in connection with the vacuum polarization
contribution, using only the energy shift gives answers in significant
disagreement with using both parts of Eq.~(\ref{eq:18}). Our result,
combining vacuum polarization with self-energy, is
\begin{equation}
R_{2p_{1/2}}(Z\alpha) \Big|_{\text{present work}}
 = (Z\alpha)^2 \biggl\{
\biggl[ {8 \over 3}\, \ln{4 \over 3} - {61 \over 18} \biggr]\!
\ln(Z\alpha)^{-2} + 6.576~03 \biggr\}.
\label{mainres}
\end{equation}
As noted earlier, we do have agreement with the logarithmic contribution
found earlier by Ivanov and Karshenboim \cite{Ivanov}. Using the
hydrogenic value of $\Gamma_0$ (including recoil through a factor
of $m_r/m_e$)
$$
\Gamma_0(2p_{1/2}) \Big|_{Z=1} = 6.264~942~3 \times 10^{8}~{\rm s}^{-1},
$$
the radiatively corrected lifetimes of the $2p_{1/2}$ state for
hydrogen are
\begin{eqnarray}
\Gamma(2p_{1/2}) &=& 6.264~927~4 \times 10^{8}~{\rm s}^{-1}
~~\text{(present work)}, \nonumber \\
\Gamma(2p_{1/2}) &=& 6.264~922~3 \times 10^{8}~{\rm s}^{-1}
~~\text{(Ref.~\cite{Ivanov})}, \nonumber \\
\Gamma(2p_{1/2}) &=& 6.264~881~2 \times 10^{8}~{\rm s}^{-1}
~~\text{(Ref.~\cite{Sok})}. \nonumber
\end{eqnarray}
As indicated earlier, Ref.~\cite{Ivanov} included only the logarithmic
term, which was in agreement with our results, so the numerical difference
shown above is due to the constant term in Eq.~(\ref{mainres}).
The difference is under 1 part per million (ppm), which corresponds
to under 1 kHz in the Lamb shift. However, there is a more significant
7 ppm difference with Ref.~\cite{Sok}, which should play a significant
role in the interpretation of that experiment. Of course, this is only
relevant if ppm precision can be reached experimentally. Issues involved
in reaching this extremely high accuracy, which we note is two orders of
magnitude greater than found in positronium \cite{orthoexp}, have been
discussed by Hinds \cite{Hinds}.

\begin{acknowledgments}
The work of J.S. was supported in part by NSF Grant No.~PHY-0097641.
The work of K.P was supported by EU Grant No.~HPRI-CT-2001-50034.
The work of K.T.C. was performed under the auspices of the U.S.
Department of Energy by Lawrence Livermore National Laboratory under
Contract No.~W-7405-Eng-48.
\end{acknowledgments}

\newpage

\begin{table*}[p]
\caption{\label{tab1}
Lowest-order one-photon decay rates to the ground state for $n=2$ states
of hydrogenic ions in atomic units. Numbers in square brackets indicate
powers of 10. The last column gives the nuclear fermi distribution
parameter $c$ in fermis. Conversion to the unit of $s^{-1}$ is through
1 a.u.\ = $4.134~137 \times 10^{16}~s^{-1}$.}
\begin{ruledtabular}
\begin{tabular}{cllll}
  \multicolumn{1}{c}{$Z$}
& \multicolumn{1}{c}{$2s_{1/2}$}
& \multicolumn{1}{c}{$2p_{1/2}$}
& \multicolumn{1}{c}{$2p_{3/2}$}
& \multicolumn{1}{c}{$c$}
\\
\colrule
  5  &  5.9038[-16]  &  9.4779[-6]  &  9.4735[-6]  &  1.8104  \\
 10  &  6.0733[-13]  &  1.5172[-4]  &  1.5144[-4]  &  2.9889  \\
 15  &  3.5262[-11]  &  7.6868[-4]  &  7.6546[-4]  &  3.2752  \\
 20  &  6.3251[-10]  &  2.4321[-3]  &  2.4140[-3]  &  3.7188  \\
 25  &  5.9680[-9]   &  5.9461[-3]  &  5.8769[-3]  &  4.0706  \\
 30  &  3.7551[-8]   &  1.2351[-2]  &  1.2144[-2]  &  4.4454  \\
 40  &  6.9521[-7]   &  3.9207[-2]  &  3.8037[-2]  &  4.9115  \\
 50  &  6.8431[-6]   &  9.6268[-2]  &  8.8114[-2]  &  5.4595  \\
 60  &  4.5463[-5]   &  2.0100[-1]  &  1.8737[-1]  &  5.8270  \\
 70  &  2.3180[-4]   &  3.7535[-1]  &  3.4051[-1]  &  6.2771  \\
 80  &  9.8091[-4]   &  6.4597[-1]  &  5.6706[-1]  &  6.6069  \\
 90  &  3.6293[-3]   &  1.0440      &  8.8110[-1]  &  6.9264  \\
100  &  1.2193[-2]   &  1.6033      &  1.2913      &  7.1717  \\
\end{tabular}
\end{ruledtabular}
\end{table*}

\begin{table*}[p]
\caption{\label{tab2}
Breakdown of contributions to $R_{2s_{1/2}}(Z\alpha)$.}
\begin{ruledtabular}
\begin{tabular}{crrrrrrrrr}
  \multicolumn{1}{c}{$Z$}
& \multicolumn{1}{c}{$Q_{V1}$}
& \multicolumn{1}{c}{$Q_{V2}$}
& \multicolumn{1}{c}{$Q_{V3}$}
& \multicolumn{1}{c}{$Q_{SL1}$}
& \multicolumn{1}{c}{$Q_{SL2}$}
& \multicolumn{1}{c}{$Q_{SR1}$}
& \multicolumn{1}{c}{$Q_{SR2}$}
& \multicolumn{1}{c}{$Q_{derb}$}
& \multicolumn{1}{c}{$R(Z\alpha)$}
\\
\colrule
 50 & -8418.269 & -68461.495 & 78211.835 & -10142.571 & 11.673 & 8593.159 & 11.810 & 192.382 & -1.476 \\
 60 & -2446.426 & -14508.277 & 17540.361 &  -4213.698 & 11.696 & 3490.708 & 11.893 & 110.828 & -2.915 \\
 70 &  -824.257 &  -4479.499 &  5587.689 &  -1990.665 & 11.725 & 1610.892 & 11.995 &  68.086 & -4.034 \\
 80 &  -305.499 &  -1593.601 &  2044.727 &  -1034.946 & 11.762 &  817.408 & 12.121 &  43.683 & -4.345 \\
 90 &  -119.499 &   -639.062 &   834.541 &   -580.832 & 11.808 &  446.644 & 12.277 &  28.790 & -5.333 \\
100 &   -47.230 &   -288.465 &   373.852 &   -347.601 & 11.867 &  259.536 & 12.474 &  19.246 & -6.321 \\
\end{tabular}
\end{ruledtabular}
\end{table*}

\begin{table*}[p]
\caption{\label{tab3}
Breakdown of contributions to $R_{2p_{1/2}}(Z\alpha)$.}
\begin{ruledtabular}
\begin{tabular}{crrrrrrrrr}
  \multicolumn{1}{c}{$Z$}
& \multicolumn{1}{c}{$Q_{V1}$}
& \multicolumn{1}{c}{$Q_{V2}$}
& \multicolumn{1}{c}{$Q_{V3}$}
& \multicolumn{1}{c}{$Q_{SL1}$}
& \multicolumn{1}{c}{$Q_{SL2}$}
& \multicolumn{1}{c}{$Q_{SR1}$}
& \multicolumn{1}{c}{$Q_{SR2}$}
& \multicolumn{1}{c}{$Q_{derb}$}
& \multicolumn{1}{c}{$R(Z\alpha)$}
\\
\colrule
  5 & -9.551 & -36.706 & 23.027 & -0.013 & 11.626 & -0.021 & 11.625 &  0.000 & -0.014 \\
 10 & -6.887 & -28.062 & 11.756 & -0.044 & 11.625 & -0.064 & 11.630 &  0.000 & -0.045 \\
 15 & -5.403 & -25.850 &  8.106 & -0.082 & 11.627 & -0.119 & 11.640 & -0.001 & -0.082 \\
 20 & -4.409 & -25.047 &  6.358 & -0.126 & 11.631 & -0.183 & 11.653 & -0.003 & -0.126 \\
 25 & -3.685 & -24.732 &  5.370 & -0.173 & 11.635 & -0.252 & 11.670 & -0.003 & -0.172 \\
 30 & -3.131 & -24.632 &  4.760 & -0.222 & 11.641 & -0.325 & 11.691 & -0.012 & -0.230 \\
 40 & -2.344 & -24.662 &  4.097 & -0.319 & 11.655 & -0.473 & 11.743 & -0.031 & -0.334 \\
 50 & -1.822 & -24.806 &  3.793 & -0.409 & 11.673 & -0.623 & 11.810 & -0.065 & -0.449 \\
 60 & -1.464 & -24.984 &  3.654 & -0.488 & 11.697 & -0.770 & 11.893 & -0.121 & -0.583 \\
 70 & -1.219 & -25.166 &  3.594 & -0.553 & 11.727 & -0.909 & 11.995 & -0.207 & -0.738 \\
 80 & -1.057 & -25.349 &  3.570 & -0.599 & 11.764 & -1.037 & 12.121 & -0.334 & -0.921 \\
 90 & -0.962 & -25.533 &  3.558 & -0.629 & 11.812 & -1.146 & 12.277 & -0.521 & -1.144 \\
100 & -0.928 & -25.726 &  3.544 & -0.637 & 11.876 & -1.229 & 12.474 & -0.800 & -1.426 \\
\end{tabular}
\end{ruledtabular}
\end{table*}

\begin{table*}[p]
\caption{\label{tab4}
Breakdown of contributions to $R_{2p_{3/2}}(Z\alpha)$.}
\begin{ruledtabular}
\begin{tabular}{crrrrrrrrr}
  \multicolumn{1}{c}{$Z$}
& \multicolumn{1}{c}{$Q_{V1}$}
& \multicolumn{1}{c}{$Q_{V2}$}
& \multicolumn{1}{c}{$Q_{V3}$}
& \multicolumn{1}{c}{$Q_{SL1}$}
& \multicolumn{1}{c}{$Q_{SL2}$}
& \multicolumn{1}{c}{$Q_{SR1}$}
& \multicolumn{1}{c}{$Q_{SR2}$}
& \multicolumn{1}{c}{$Q_{derb}$}
& \multicolumn{1}{c}{$R(Z\alpha)$}
\\
\colrule
   5 & -9.551 & -36.484 & 22.775 &  0.001 & 11.626 & -0.006 & 11.625 & 0.000 & -0.014 \\
  10 & -6.888 & -27.653 & 11.254 &  0.000 & 11.625 & -0.014 & 11.630 & 0.000 & -0.045 \\
  15 & -5.406 & -25.318 &  7.397 & -0.001 & 11.627 & -0.023 & 11.640 & 0.001 & -0.082 \\
  20 & -4.413 & -24.381 &  5.418 & -0.002 & 11.630 & -0.033 & 11.653 & 0.004 & -0.123 \\
  25 & -3.690 & -23.946 &  4.202 & -0.004 & 11.635 & -0.042 & 11.670 & 0.008 & -0.166 \\
  30 & -3.138 & -23.732 &  3.368 & -0.008 & 11.640 & -0.050 & 11.691 & 0.014 & -0.215 \\
  40 & -2.352 & -23.572 &  2.278 & -0.022 & 11.656 & -0.063 & 11.743 & 0.037 & -0.295 \\
  50 & -1.830 & -23.555 &  1.579 & -0.047 & 11.675 & -0.075 & 11.810 & 0.082 & -0.361 \\
  60 & -1.470 & -23.594 &  1.082 & -0.087 & 11.698 & -0.087 & 11.893 & 0.144 & -0.421 \\
  70 & -1.218 & -23.661 &  0.710 & -0.149 & 11.726 & -0.102 & 11.995 & 0.242 & -0.457 \\
  80 & -1.043 & -23.751 &  0.424 & -0.237 & 11.758 & -0.119 & 12.121 & 0.383 & -0.464 \\
  90 & -0.926 & -23.863 &  0.204 & -0.360 & 11.794 & -0.140 & 12.277 & 0.585 & -0.429 \\
100  & -0.856 & -24.006 &  0.042 & -0.527 & 11.833 & -0.160 & 12.474 & 0.877 & -0.323 \\
\end{tabular}
\end{ruledtabular}
\end{table*}

\end{document}